\documentclass[12pt,a4paper,english,superscriptaddress,aps,nofootinbib]{revtex4}
\usepackage[utf8]{inputenc}
\usepackage[T1]{fontenc}
\usepackage{amsmath,amssymb,graphicx}
\makeatletter
\usepackage{babel}
\usepackage[active]{srcltx}
\usepackage{graphicx,color}
\usepackage{changebar}
\usepackage{hyperref}
\usepackage[T1]{fontenc}
\usepackage{esint}
\usepackage{multirow}
\usepackage{dsfont}
\usepackage{ae}
\usepackage{amsmath}
\usepackage{braket}
\usepackage{mathtools}
\usepackage{slashed}
\usepackage{empheq}
\usepackage{multirow}
\usepackage{graphicx}
\usepackage{amsfonts}
\usepackage{amsmath}
\newcommand{\be}{\begin{equation}}
	\newcommand{\ee}{\end{equation}}
\newcommand{\bea}{\begin{eqnarray}}
	\newcommand{\eea}{\end{eqnarray}}

\begin{document}

\title{Topological solitons in the sigma-cuscuton model}

\author{F. C. E. Lima}
\email{cleiton.estevao@fisica.ufc.br}
\affiliation{Departamento do F\'{i}sica, Universidade Federal do Cear\'{a}, 60455-760, Fortaleza, CE, Brazil.}

\author{C. A. S. Almeida}
\email{carlos@fisica.ufc.br}
\affiliation{Departamento do F\'{i}sica, Universidade Federal do Cear\'{a}, 60455-760, Fortaleza, CE, Brazil.}
\affiliation{Institute of Cosmology, Department of Physics and Astronomy, Tufts University, Medford, Massachusetts 02155, USA.}

\begin{abstract}\vspace{0.4cm}
\noindent \textbf{Abstract:} Building a multi-field theory with canonical and non-canonical contributions, one studies the topological solitons of the O(3)-sigma model. We propose a model constituted by the O(3)-sigma field, the cuscuton-like neutral scalar field, and Maxwell's field. We investigate BPS properties considering a theory without interaction. One performs this study by adopting the first-order formalism in a model with contribution non-canonical. Thus, these contributions will preserve the spontaneous symmetry breaking of the system. Concurrently, a non-minimal coupling between the sigma and the Maxwell field is assumed. In this scenario, interesting results arise, i.e., one notes that the solitons have an internal structure and ring-like profile. Furthermore, one observes that the ring-like configurations that emerge are directly related to the contribution of the cuscuton-like term.

\end{abstract}
\maketitle

\thispagestyle{empty}


\section{Introduction}

The physics of planar structures describes interesting properties \cite{Bais}, e. g., charge fractioning \cite{Feldman,Cherman} and fractional statistics \cite{Arovas}. Furthermore, in analyzing planar systems, several interesting features arise due to the correspondence between particles and their duals. One of these correspondences is the particle-vortex duality \cite{Karch,Murugan,Metlitski}. In the planar world, vortices constitute an important class of structures. The importance of these structures is due to their relevant applications, as we can see in Refs. \cite{Lima1,Lima2,Lima3,Lima4}. A notably interesting application appears in condensed matter physics, where these structures appear in the phenomena description of superconductivity \cite{Abrikosov,Davis1,Davis2}.

In general, one can understand the vortices as structures that arise in three-dimensional spacetime, i. e., $(2+1)$D \cite{Casana2,Casana3,Casana4,Edery1,Edery2}. In field theory, pioneers in the study of vortex structures were Nielsen and Olesen \cite{Nielsen}. In the seminal paper: {\it Vortex-line models in dual strings}, the authors show the vortex solutions of an action constructed with a complex scalar field minimally coupled to a gauge field with symmetry $U(1)$ \cite{Nielsen}. After Nielsen and Olesen's proposal, several papers emerged discussing topological \cite{Weinberg,Hong} and non-topological \cite{LeeP,Arthur,Kimm} structures.

Only in 1991, Stern \cite{Stern} proposed for the first time the study of a theory non-minimally coupled to the gauge field. Using a three-dimensional spacetime, Stern seeks to describe point particles with no spin degree of freedom that carries an appropriate magnetic momentum. Stern's work motivated several researchers who later proposed papers on non-minimal models, e. g., vortices non-minimally coupled to the gauge field \cite{Lima3,Torres,PKGhosh,CA,SGhosh}. To be specific, in Ref. \cite{Cavalcante}, the authors investigate BPS vortex solutions for a specific interaction using an $O(3)$-sigma model non-minimally coupled to a Maxwell-Chern-Simons field. Besides, the BPS properties of sigma model vortices were also studied using a non-minimal coupling and a multi-field approach \cite{Lima3}. Motivated by these applications, a natural questioning arises: How are vortex structures modified in a non-minimum theory constituted by non-canonical multi-fields? Throughout this work, we will expose the answer to this question.

In this research article, we use the non-linear O(3)-sigma model. Briefly, the non-linear O(3)-sigma model consists of three real scalar fields \cite{Rajaraman}, i. e., $\Phi(\textbf{r},t)\equiv\{\phi_i(\textbf{r},t),\, i=1,2,3\}$ with the constraint
\begin{align}\label{vin}
    \Phi\cdot\Phi=\sum_{i=1}^{3}\phi_i\phi^i=1.
\end{align}
Respecting this constraint, the dynamics of the O(3)-sigma field, i. e., of the field $\Phi$ is governed by the following Lagrangian
\begin{align}
    \mathcal{L}=\frac{1}{2}\partial_\mu\Phi\cdot\partial^\mu\Phi.
\end{align}
Thus, one describes the sigma model as a vector of fields in its internal space, i. e., a three-dimensional field space \cite{Rajaraman,Ghosh1,Ghosh2,Schroers1,Schroers2}. In 1960, Gell-Mann and L\'{e}vy were the first to propose this model \cite{Gellmann}. At the time, the purpose was to describe the Goldberger and Treiman formula for the rate of decay of the charged pion using a strong interaction proposed by Schwinger \cite{Schwinger} and a weak current formulated by Polkinghorne \cite{Polkinghorne}. After the work of Gell-Mann and L\'{e}vy, several papers considered the non-linear sigma model in their analysis. For example, using the O(3)-sigma model, photons emerging was investigated in Ref. \cite{Motrunich}. Furthermore, the solitons stability and Lorentz violation were studied, respectively, in Refs. \cite{Leese} and \cite{Messias}.

Not far from the non-linear sigma model, some authors have proposed so-called multi-field models \cite{Lima3}. These models play an important role in inflationary theories \cite{Langlois1}. That is because the theoretical results of the multi-field theories agree with the phenomenological measurements \cite{Langlois1,Langlois2,Bean,Trotta,Keskitalo}. Thus, that motivates us to study the topological structures derived from this kind of theory. Indeed, one can find some research articles in the literature discussing aspects of structures in multi-field theories, e. g., see Refs. \cite{Oles,Liu}. However, as far as we know, no study was performed discussing the vortex structures considering an O(3)-sigma and other non-canonical fields.

In particular, in this work, in addition to the dynamic term of the sigma model, we will use a cuscuton-like non-canonical real scalar field. Afshordi, Chung, and Geshnizjani announce the cuscuton model in the paper: {\it A causal field theory with an infinite speed of sound} \cite{Afshordi}. In this theory, the cuscuton dynamics arise from the degenerate Hamiltonian symplectic structure description in the cosmologically homogeneous limit \cite{Afshordi}. In this case, the cuscuton theory becomes homogeneous when the metric is locally Minkowski \cite{Afshordi,Afshordi2,Afshordi3}. An interesting feature of the cuscuton field is that it does not contribute to the equation of motion at the stationary limit. Thus, one can interpret it as a non-dynamic auxiliary field that follows the dynamics of the fields to which it couples.

Naturally, together with these applications and motivations arise some questioning. For example, is it possible to obtain a vortex line in an O(3)-sigma model coupled to a non-canonical field? How do the non-canonical term and multi-field influence the structure of O(3)-sigma vortices? These are relevant questions that motivate our study. Thus, considering a sigma-cuscuton model, we hope to answer these questions throughout this research article.

We organized our work as follows: In Sec. II, the BPS vortices are analyzed. In Sec. III, we implement spherical symmetry in the target space of the O(3)-sigma model. Posteriorly, in Sec. IV, topological vortex solutions are displayed. To finalize, in Sec. V, our findings are announced.

\section{Non-minimal BPS vortex}

In this work, we will study for the first time the vortex solutions of the non-minimal O(3)-sigma model coupled to the gauge and the cuscuton-like scalar fields. This model can be, in principle, useful to describe soliton's crystals of non-canonical theories with contributions of anomalous magnetic dipole momentum (AMDM). Indeed, this is possible because O(3)-sigma models can characterize gauged soliton crystals and allow the appearance of electroma- gnetic structures with nonzero topological charges. For more details, see Ref. \cite{Canfora}. Further- more, in our study, we will adopt a non-minimal model, i.e., we will consider that the structures have a contribution from AMDM. One motivates this because the physical properties of the particles are modified when subject to AMDM \cite{Chaudhuri}. Thus, in our theory, the AMDM can, in principle, change the energy from the structures, making them more or less energetic. Besides, the AMDM can induce the emergence of internal and ringlike structures \cite{Lima3}. Finally, we will consider a cuscuton-like non-canonical contribution to our theory. That is because these contributions, widely used in cosmological models \cite{Bartolo,Iyonaga}, can provide us with an answer about the influence of non-canonical theories on the physical properties of structures, in our case, the magnetic vortex. Thus, let us start our study by considering a three-dimensional model, i.e., a spacetime with $(2+1)$D. In this scenario, the Lagrangian density of our theory is
\begin{align}\label{Lag}
    \mathcal{L}=\frac{1}{2}\nabla_{\mu}\Phi\cdot\nabla^{\mu}\Phi+\eta\sqrt{\vert \partial_\mu\psi\partial^\mu\psi\vert}+\frac{1}{2}\partial_\mu\psi\partial^\mu\psi-\frac{1}{4}F_{\mu\nu}F^{\mu\nu}-\mathcal{V}(\phi_3,\psi).
\end{align}
Here, $\Phi$ is a triplet of scalar fields subject to the constraint $\Phi\cdot\Phi=1$. Meanwhile, $\psi$ is a real scalar field, $F_{\mu\nu}=\partial_\mu A_\nu-\partial_\nu A_\mu$ is the electromagnetic tensor and $\mathcal{V}( \phi_3,\psi)$ is the theory interaction. Furthermore, the term $\eta\sqrt{\vert\partial_\mu\psi\,\partial^\mu\psi\vert}$ is known as the cuscuton term. This term describes non-canonical theories  \cite{Afshordi,Afshordi2,Afshordi3}. Indeed, the term cuscuton appears for the first time as an alternative to describe dark matter and their contribution to the action lacks a dynamic degree of freedom \cite{Lima2,Afshordi2}. In its etymology, the word \textit{cuscuton} originates in Latin and describes a parasitic plant, namely, the Cuscuta. Based on this, we call our theory of sigma-cuscuton-like model.

As discussed in Ref. \cite{Lima3}, one defines the usual covariant derivative as
\begin{align}
    \label{CovariantD0}
    D_\mu\Phi=\partial_\mu\Phi+eA_\mu\,(\hat{n}_3\times\Phi).
\end{align}
Meanwhile, the non-minimal covariant derivative is 
\begin{align}\label{CovariantD}
    \nabla_\mu\Phi=\partial_\mu\Phi+\bigg(eA_\mu+\frac{g}{2}\varepsilon_{\mu\nu\lambda}F^{\nu\lambda}\bigg)\,\hat{n}_3\times \Phi.
\end{align}
Let us study the non-minimal theory, i.e., vortex configurations with an anomalous contribution of magnetic momentum. We introduce the anomalous magnetic momentum contribution using the coupling $\frac{g}{2}\varepsilon_{\mu\nu\lambda}F^{\nu\lambda}$ in the covariant derivative, i. e., a coupling between the gauge field and the matter field. One can find the non-minimal coupling applied in investigations of the properties of BPS solitons, e. g., see Refs. \cite{Torres,PKGhosh,CA,SGhosh}.

To carry out our study, allow us to consider a flat spacetime with a metric signature such as $\eta_{\mu\nu}=$ det$(-,+,+)$. Moreover, one defines the gauge field as
\begin{align}\label{GaugeEquation}
    j^\nu=\partial_\lambda[g\varepsilon_{\mu\lambda\nu}(\Phi\times\nabla^\mu\Phi)\cdot\hat{n}_3-F^{\lambda\nu}],
\end{align}
where $j^\nu=e(\Phi\times\nabla^\nu\Phi)\cdot\hat{n}_3$ and $\textbf{J}^\nu=-j^\nu\cdot\hat{n}_3$.

By inspection of Gauss' law, i. e., the component $\nu=0$ of Eq. (\ref{GaugeEquation}), we can assume $A_0=0$. In this case, the structures that arise in this theory will be purely magnetic.

Investigating the equation of motion, one obtains the matter field equation, namely,
\begin{align}
    \nabla^\mu\nabla_\mu\Phi=-\mathcal{V}_\Phi,
\end{align}
with $\mathcal{V}_\Phi=\frac{\partial\mathcal{V}}{\partial \Phi}$.

Meanwhile, the real scalar field equation is 
\begin{align}
    \partial_\mu\bigg[\partial^\mu\psi+\eta\frac{\partial^\mu\psi}{\sqrt{\vert\partial_\nu\psi\,\partial^\nu\psi\vert}}\bigg]=-\mathcal{V}_\psi,
\end{align}
with $\mathcal{V}_\psi=\frac{\partial\mathcal{V}}{\partial \psi}$.

Before investigating the topological vortex structures, allow us to present some useful preliminary definitions for our study. In this work, we will study spherically symmetric vortex structures. Thereby, let us assume the ansatz proposed by Schroers in Ref. \cite{Schroers1}, i. e.,
\begin{align}\label{ansatz1}
    \Phi(r, \theta)=\begin{pmatrix}
    \sin f(r)\cos N\theta\\
    \sin f(r)\sin N\theta\\
    \cos f(r)
    \end{pmatrix}.
\end{align}
This ansatz is necessary for the $\Phi$ field to respect the O(3)-sigma model constraint, i. e., $\Phi\cdot\Phi=1$. It is interesting to mention that this ansatz was used widely in other works, e. g., see Refs. \cite{Lima5,Lima6}.

On the other hand, as suggested in Refs. \cite{Lima3,Casana5}, the real scalar field is
\begin{align}
    \psi=\psi(r).
\end{align}

To study the vortex configurations, we use the ansatz proposed in Refs. \cite{Schroers1,PKGhosh}, i. e.,
\begin{align}\label{ansatz3}
    \textbf{A}(r)=-\frac{Na(r)}{er}\hat{\textbf{e}}_{\theta},
\end{align}
where $N$ is the winding number. Being the magnetic field defined as $B=-F_{12}$, one obtains that the magnetic field in terms of $a(r)$ is
\begin{align}\label{MagneticF}
    B\equiv -F_{12}=\frac{Na'(r)}{er}.
\end{align}

Considering the planar nature of the vortex, we define the magnetic flux as\footnote{One defines the magnetic flux with a negative sign to obtain topological vortices with flux positive-defined consistent with the topological conditions of the gauge field.}
\begin{align}\label{PP}
    \phi_{flux}=-\int_{0}^{2\pi}\int_{0}^\infty\frac{Na'(r)}{er}rdrd\theta.
\end{align}

Once we know the preliminary definitions of the sigma, cuscuton-like scalar, and gauge fields, we can investigate the soliton-like configurations that describe topological vortices. To study these structures is necessary to investigate the energy of the system. To perform this analysis, let us construct the energy-momentum tensor and examine the $T_{00}$ component of the energy-momentum tensor. The integration from the $T_{00}$ component in the overall space gives us the energy of the structures. Performing this analysis in terms of field variables $a(r)$, $\psi(r)$, and $f(r)$, leads us to
\begin{align}\label{energy1}\nonumber
    \mathrm{E}=&\int\, d^2x\,\bigg[\frac{1}{2}\bigg(f'(r)\mp\frac{N}{r}[a(r)-1]\sin f(r)\bigg)^2+\frac{1}{2}\bigg(\psi'(r)\mp\frac{W_\psi}{r}\bigg)^2+\frac{1}{2}(B\mp\sqrt{2\mathcal{U}})^2+\\
    +&\eta\psi'(r)\pm\frac{N[a(r)-1]}{r}f'(r)\sin f(r)\pm\frac{Na'(r)}{er}\sqrt{2\mathcal{U}}\pm \frac{W_\psi \psi'}{r}-\frac{W_\psi^2}{2r}+\mathcal{V}-\mathcal{U}\bigg].
\end{align}
Here, we implement in the energy two interactions, i. e., $\mathcal{W}=\mathcal{W}[\psi(x_i);\, x_i]$ and $\mathcal{U}=\mathcal{U}[\phi_3(x_i);\, x_i ]$ with $\mathcal{W}_\psi=\frac{\partial \mathcal{W}}{\partial\psi}$. In general, one implements the superpotential functions $\mathcal{W}$ and $\mathcal{U}$ to obtain a first-order formalism of the theory. Indeed, these superpotentials play a relevant role, i. e., these functions relate with the potential $\mathcal{V}$ at the saturation limit of the energy \cite{Vachaspati}. Thus, it allows the energy saturation limit to obtain first-order equations of motion \cite{Vachaspati}, which is quite suitable for our purpose. Furthermore, one assumes that $\psi=\psi(r)$ due to the vortex symmetry, i. e., a function purely of the radial variable.

Analyzing the energy (\ref{energy1}), one notes that the static field configurations have energy bounded from below. Therefore, at the energy saturation limit, one obtains
\begin{align}\label{BPS1}
    f'(r)=\pm\frac{N}{r}[(a(r)-1)\sin f(r)], \, \, \, \, \, \, 
    a'(r)=\pm\frac{r}{N}\sqrt{\mathcal{U}} \, \, \, \, \, \, \, \text{and} \, \, \, \, \, \, \psi'(r)=\pm\frac{W_\psi}{r}.
\end{align}
Note that the first two first-order equations of the expression (\ref{BPS1}) are known as the Bogomol'nyi equations (or BPS equation) that describe the vortices of the O(3)-sigma model. On the other hand, the expression $\psi'(r)=\frac{W_\psi}{r}$ is the BPS equation for the scalar field without the contribution of the non-canonical term (the cuscuton contribution). As a matter of fact, in the stationary case, the dynamics derived from the cuscuton term do not contribute to the equation of motion. That occurs because when we consider the case of the cuscuton-like scalar field $\psi=\psi(r_1)\equiv\psi(r)$, the contribution of the cuscuton-like term in the equation of motion is 
\begin{align}
    \partial_\mu\bigg[\frac{\partial \mathcal{L}_{cusc}}{\partial(\partial_\mu\psi)}\bigg]=\bigg(\frac{\partial\mathcal{L}_{cusc}}{\partial \psi'}\bigg)'=\eta\bigg(\frac{\partial\vert\psi'\vert}{\partial\psi'}\bigg)',
\end{align}
which disappears, except in the singular case, i. e., $\psi'=0$. However, this singularity is removable. Therefore, one can assign the value zero to the contribution of the cuscuton-like to the equation of motion. Thus, pure contributions from the cuscuton term yield only a trivial contribution to the equations of motion, regardless of the shape of the potential. So, we hope that the first-order motion equation for the $\psi$ field is simply the BPS equation for the $\psi$ field without the cuscuton contributions.


To obtain the BPS properties, we assume that the interaction is
\begin{align}
    \mathcal{V}=\mathcal{U}+\frac{W_{\psi}^{2}}{2r^2}\mp\eta\frac{W_\psi}{r}.
\end{align}
Perceive that the last term in the potential is the contribution of the non-canonical term. Thus, the cuscuton-like term in the BPS limit plays the role of we call impurity. This word is applied to characterize terms in the action that do not change the equations of motion but can change the soliton profile \cite{Adam}. In truth, we can find theories with impurity in some works. For example, the impurities appear in the studies of the self-dual configuration solubility \cite{Adam}, CP$(2)$ vortex solutions \cite{Casana}.

Therefore, in the energy saturation limit, one obtains
\begin{align}\label{EnergyBPS}
    \mathrm{E}_{\text{BPS}}=\pm&\int\,\frac{1}{r}\frac{d}{dr}[N(a(r)-1)\cos f(r)+W]\, d^2x
\end{align}
Furthermore, it is important to mention that the integrand from Eq. (\ref{EnergyBPS}) is the BPS energy density. Therefore, the BPS energy density is
\begin{align}
    \label{BPSED}
    \mathcal{E}_{\text{BPS}}=\pm\frac{1}{r}\frac{d}{dr}[N(a(r)-1)\cos f(r)+W].
\end{align}

Furthermore, we highlight that the contribution of the topological sector of the sigma field on the BPS energy is
\begin{align}
    \mathrm{E}_{\text{BPS}}^{(\sigma)}=\pm&\int\,\frac{1}{r}\frac{d}{dr}[N(a(r)-1)\cos f(r)]\, d^2x.
\end{align}
On the other hand, the contribution of the topological sector of the field $\psi$ is
\begin{align}
    \mathrm{E}_{\text{BPS}}^{(\psi)}=\pm&\int\,\frac{1}{r}\frac{dW}{dr}\, d^2x,
\end{align}
so that the full BPS energy (\ref{EnergyBPS}) is
\begin{align}    \mathrm{E}_{\text{BPS}}=\mathrm{E}_{\text{BPS}}^{(\sigma)}+\mathrm{E}_{\text{BPS}}^{(\psi)}.
\end{align}

\section{Topological boundary conditions and vortex's properties}

For the study of the magnetic vortices described by the Lagrangian density (\ref{Lag}), it is necessary to assume some topological boundary conditions. Thus, we adopt the following conditions:
\begin{align}\label{top1}
    \psi(r\to -\infty)=\mp1, \hspace{1cm} \psi(r\to\infty)=\pm1,
\end{align}
\begin{align}\label{top2}
    &f(r\to 0)=0, \hspace{1cm} f(r\to \infty)=\pi,
\end{align}
and
\begin{align}
    \label{top3}
    &a(r\to 0)=0, \hspace{1cm} a(r\to \infty)=-\beta.
\end{align}
Here $\beta\in\mathds{Z}^{+}$.

The magnetic field (\ref{MagneticF}) is responsible for arising of a magnetic flux that emerges from the vortex. Thus, considering the topological condition (\ref{top3}) the magnetic flux (\ref{PP}) will be
\begin{align}\label{Mflux}
    \phi_{\text{flux}}=\frac{2\pi N}{e}[a(0)-a(\infty)],
\end{align}
i. e.,
\begin{align}
    \phi_{\text{flux}}=\frac{2\pi\beta N}{e},
\end{align}
where $N\in\mathds{Z}$.

Furthermore, topological conditions (\ref{top1}-\ref{top3}) reduce the BPS energy (\ref{EnergyBPS}) of the structures to
\begin{align}
    \mathrm{E}_{\text{BPS}}=\mp (2\pi M+\Delta W),
\end{align}
where $M\in\mathds{Z}$ is equal to $N(2+\beta)$. Besides, $\Delta W=W(r\to\infty)-W(r\to -\infty)$.

\section{Vortex solutions without interaction}

\subsection{The scalar field solutions}

To study the topological structures, we adopt a $\psi^4$-like theory. To this, we assume the superpotential
\begin{align}\label{SPW}
    W[\psi(r)]=\alpha\psi(r)\bigg[1-\frac{1}{3}\psi(r)^2\bigg].
\end{align}
To avoid carrying too many constants in our theory, let us assume $\eta=\alpha$. This choice is interesting, and its motivations are purely physical. That is because by adopting $\eta=\alpha$, the cuscuton-like non-canonical term will influence the symmetry breaking of each topological sector in the limit $\mathcal{V}\to 0$. Furthermore, it is essential to highlight that the choice $\eta=\alpha$ does not eliminate any specific solutions in the adopted regime.

The superpotential (\ref{SPW}) describes a $\phi^4$-like interaction. Therefore, when considering this superpotential, we are ensuring that spontaneous symmetry breaking occurs. This spontaneous symmetry breaking will be responsible for the arising of structures in the topological sector of $\psi$ \cite{Vachaspati}.

Now, using the superpotential (\ref{SPW}) the first-order equation of $\psi(r)$ is
\begin{align}\label{PsiE}
    \psi'(r)=\pm\frac{\alpha}{r}[1-\psi(r)^2],
\end{align}
lead us to
\begin{align}\label{tanhln}
    \psi(r)=\pm\tanh[\text{ln}(r^\alpha)].
\end{align}
The solutions of the $\psi$ field are called kink-like (positive sign) and antikink-like (negative sign) solutions\footnote{One can analytical proof this solution. We show this solution in Appendix A.}. In Fig. \ref{fig1}, we display the kink-like and antikink-like solutions that describe the field $\psi$.

\begin{figure}[!ht]
    \centering
    \includegraphics[height=6.5cm,width=7.5cm]{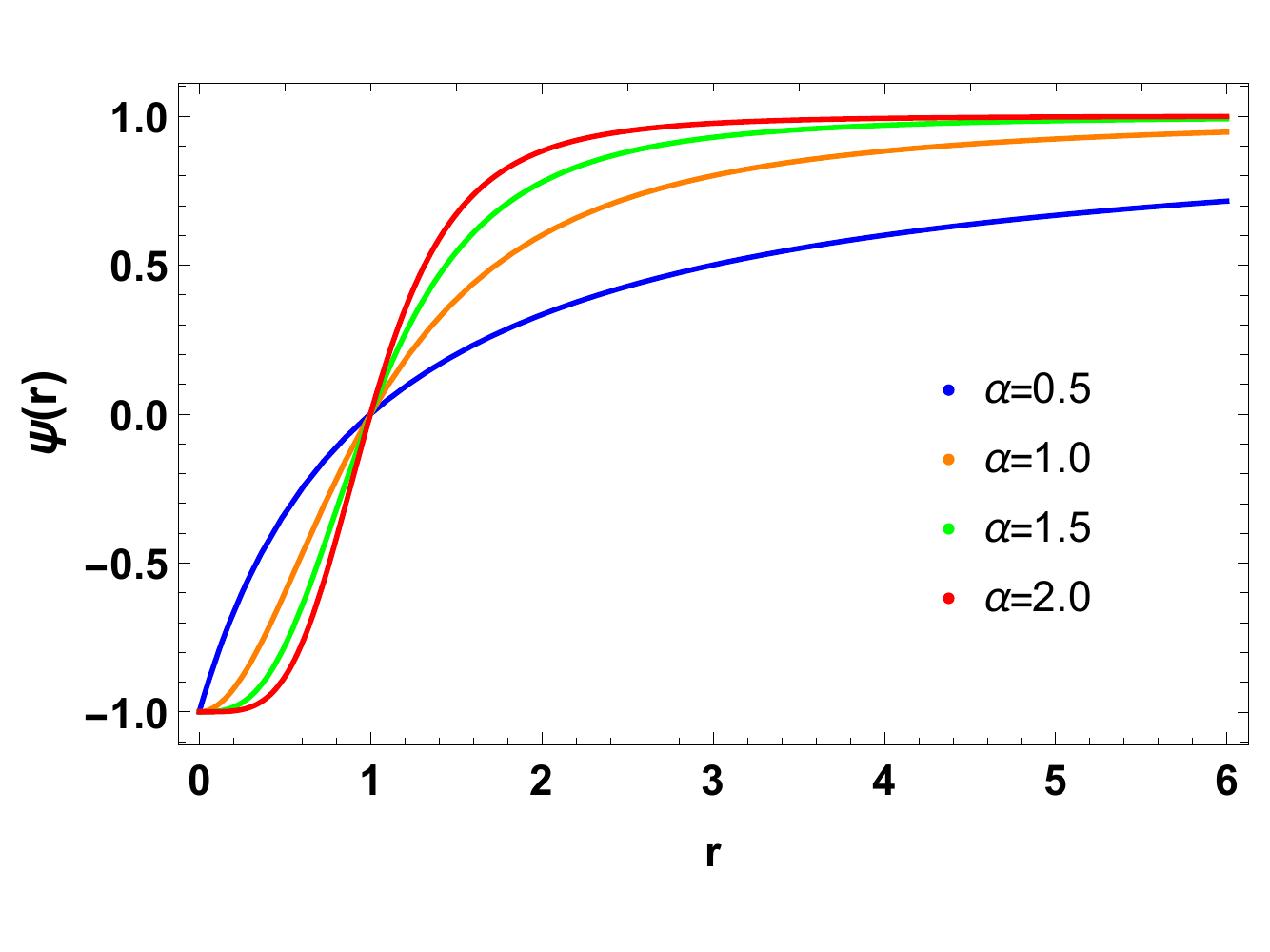}
    \includegraphics[height=6.5cm,width=7.5cm]{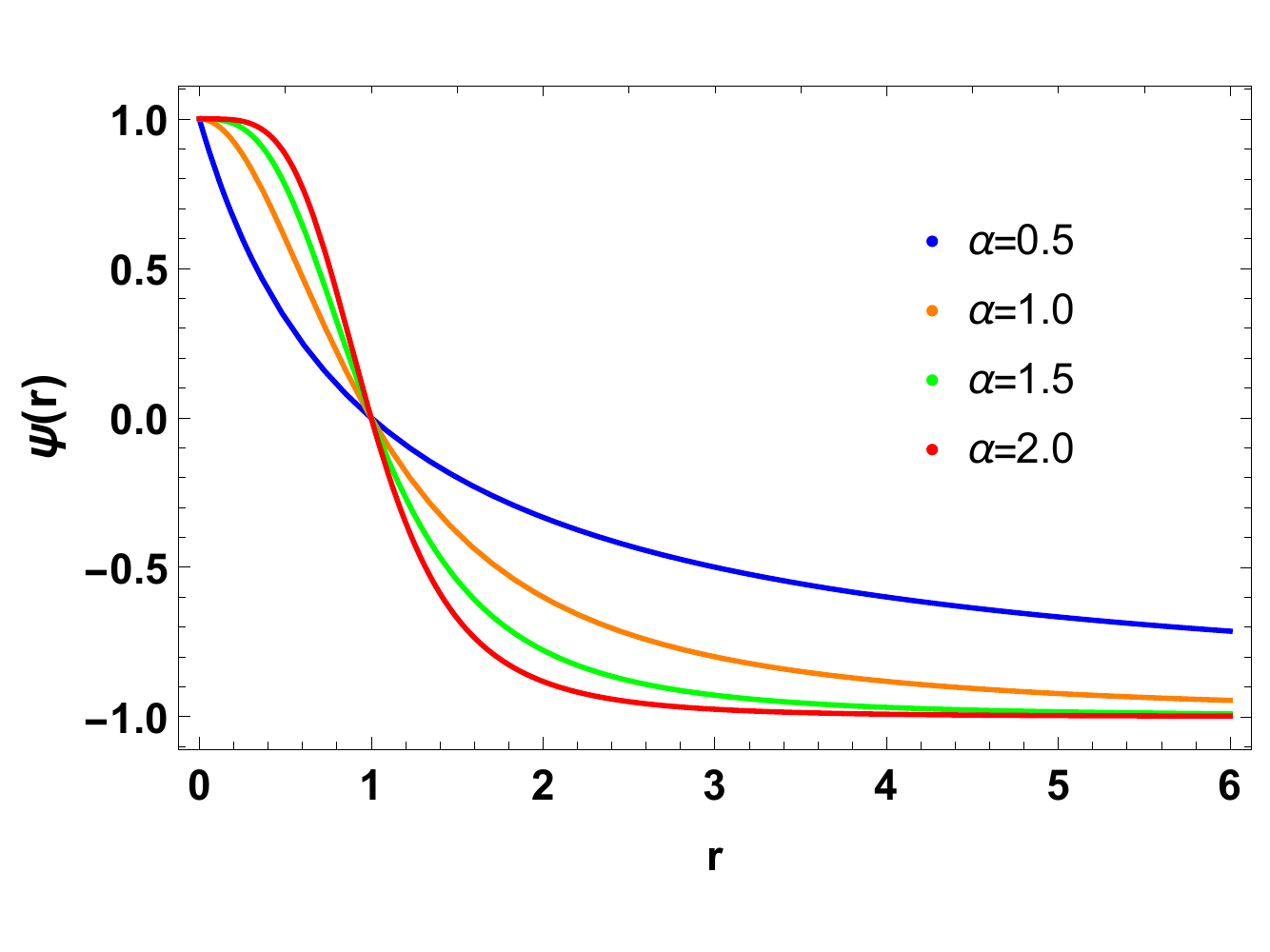}\\
    \vspace{-1cm}
    \begin{center}
       \hspace{0.5cm} (a) \hspace{7cm} (b)
    \end{center}
    \vspace{-0.5cm}
    \caption{Solutions of $\psi(r)$. (a) kink-like configuration. (b) Antikink-like configuration.}
    \label{fig1}
\end{figure}

\subsection{The theory without interaction}

Allow us to consider a theory without interaction to investigate the vortex solutions of the non-minimal sigma-cuscuton model. Thus, to obtain a model without interaction, the constraint
\begin{align}\label{UP}
    \mathcal{U}=-\frac{W_\psi^2}{2r^2}\pm\alpha\frac{W_\psi}{r}
\end{align}
must be satisfied. In this case, the interaction of the theory [see the Lagrangian (\ref{Lag})] is null, i.e., $\mathcal{V}=0$. 

\subsection{The vortex solutions without interaction}

Let us now investigate the possibility of the existence of a topological solitons\footnote{Understand the topological solutions as configurations that respect the conditions (\ref{top1}-\ref{top3}).} described by the system of equations
\begin{align}\label{B1}
    f'(r)=\pm\frac{N}{r}[(a(r)-1)\sin f(r)] \, \, \, \, \, \text{and} \, \, \, \, \, \,  a'(r)=\pm\frac{r}{N}\sqrt{\pm\alpha\frac{W_\psi}{r}-\frac{W_\psi^2}{2r^2}}.
\end{align}

\begin{figure}[ht!]
    \centering
    \includegraphics[height=6.5cm,width=7.5cm]{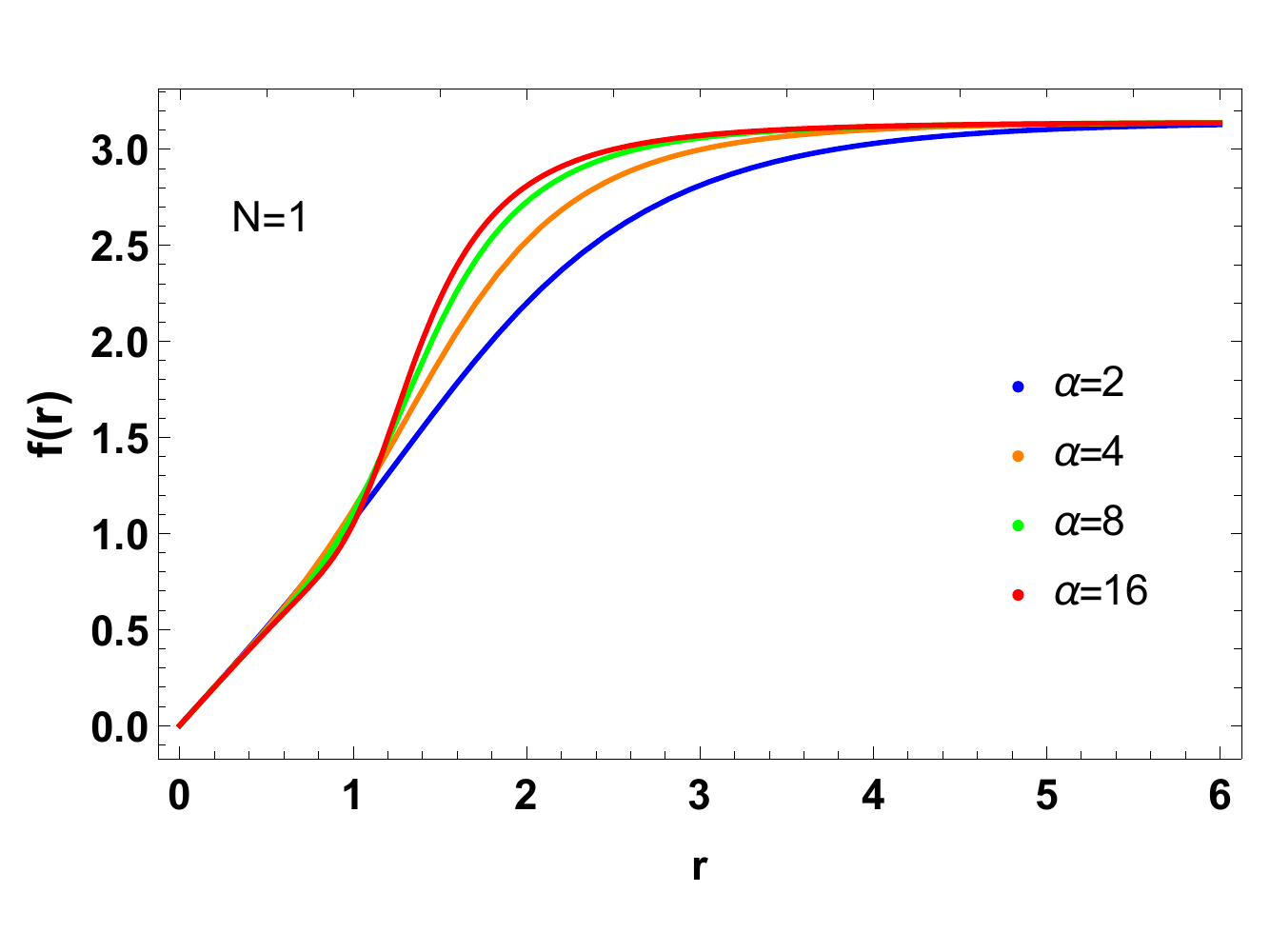}
    \includegraphics[height=6.5cm,width=7.5cm]{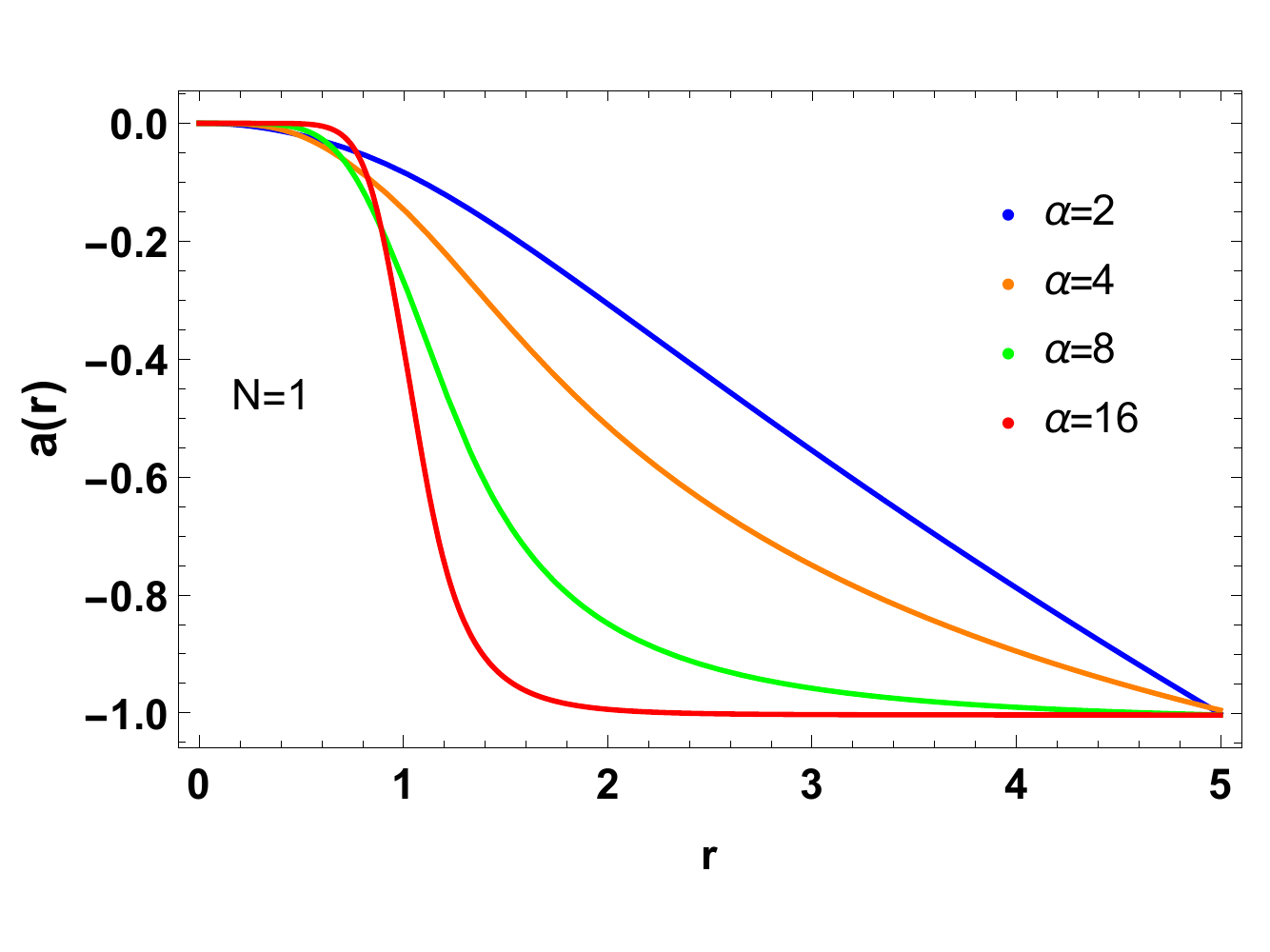}\vspace{-1cm}
    \begin{center}
        \vspace{0.5cm} (a) \hspace{7cm} (b)
    \end{center}
    \vspace{-0.5cm}
    \caption{(a) Solution of the field variable of the O(3)-sigma model. (b) Solution of the gauge field. In both plots, the dotted line is the curve when $\alpha=1$, while the other curves correspond to $\alpha=2,4,8,16$ and $32$.}
    \label{fig2}
\end{figure}

Considering the topological boundary conditions (\ref{top2}) and (\ref{top3}), let us investigate the vortex solutions produced by Eqs. (\ref{B1}). To study these solutions, we will use the numerical interpolation method\footnote{To obtain the numerical solutions of the sigma sector, we build a numerical interpolation routine with the topological boundary conditions (\ref{top1}-\ref{top3}). Furthermore, we require that at the origin of the vortex the fields $f(r)$ and $a(r)$ are null, respecting all system constraints. That allows us to linearize numerically the field equations near to the origin. Thus, we obtain that near the origin $f(r)\approx b r^{n}$ where $n$ is the winding number, and $\nu$ is a positive constant. This implies that in the vicinity from the origin $f'(r)\approx b nr^{n-1}$, so that for $n=1$, $f'(0)=b$ and for $n>1$ , $f'(0)=0$. Therefore, for $n=1$, $f(r)$ has a positive slope at $r\approx 0$ for any winding number. Furthermore, numerically we linearize $a(r)$, so that $a(r)=c r^{2}$ where $c$ is some negative constant. This implies that $a'(0)\approx 0$. Thus, we build an interpolation routine with well-defined structures at the origin that respect the topological conditions (\ref{top2}) and (\ref{top3}).}. Thus, in Fig. \ref{fig2}, the numerical solutions are displayed. Fig. \ref{fig2}(a) corresponds to the matter field solutions of the topological sector for the $\Phi$ field. On the other hand, Fig. \ref{fig2}(b) corresponds to the topological solutions of the gauge field.

\begin{figure}[ht!]
    \centering
    \includegraphics[height=3.2cm,width=4.2cm]{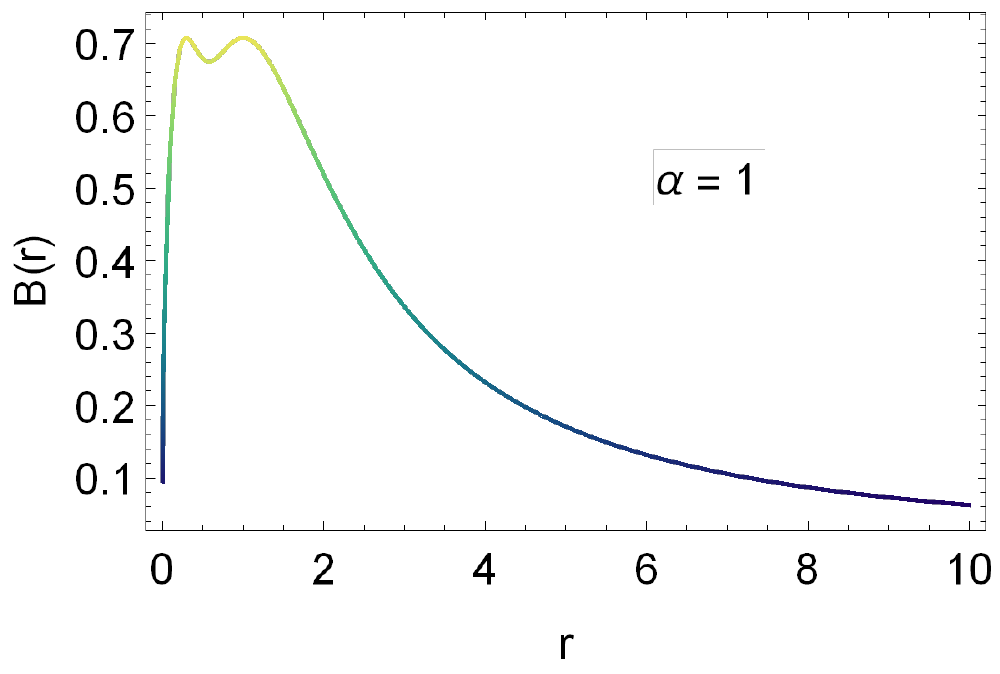}
    \includegraphics[height=3cm,width=3cm]{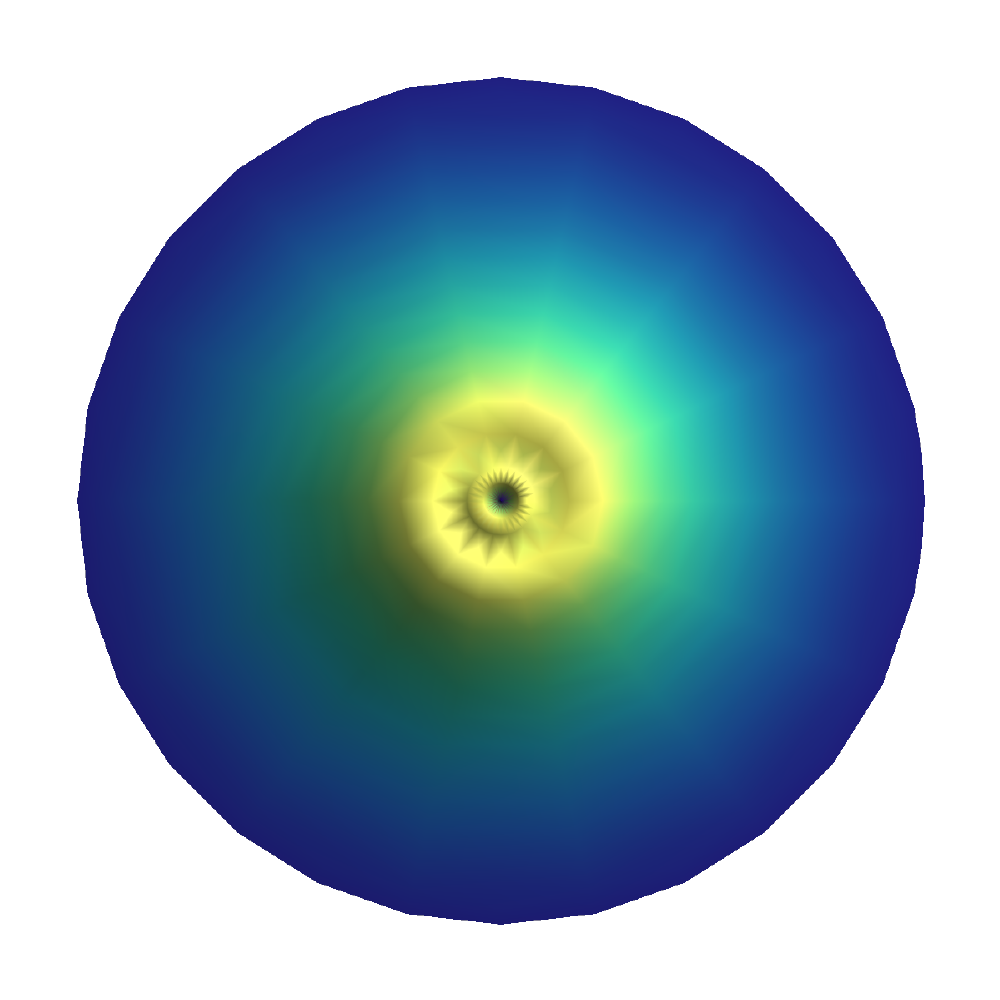}\\
    \includegraphics[height=3.2cm,width=4.2cm]{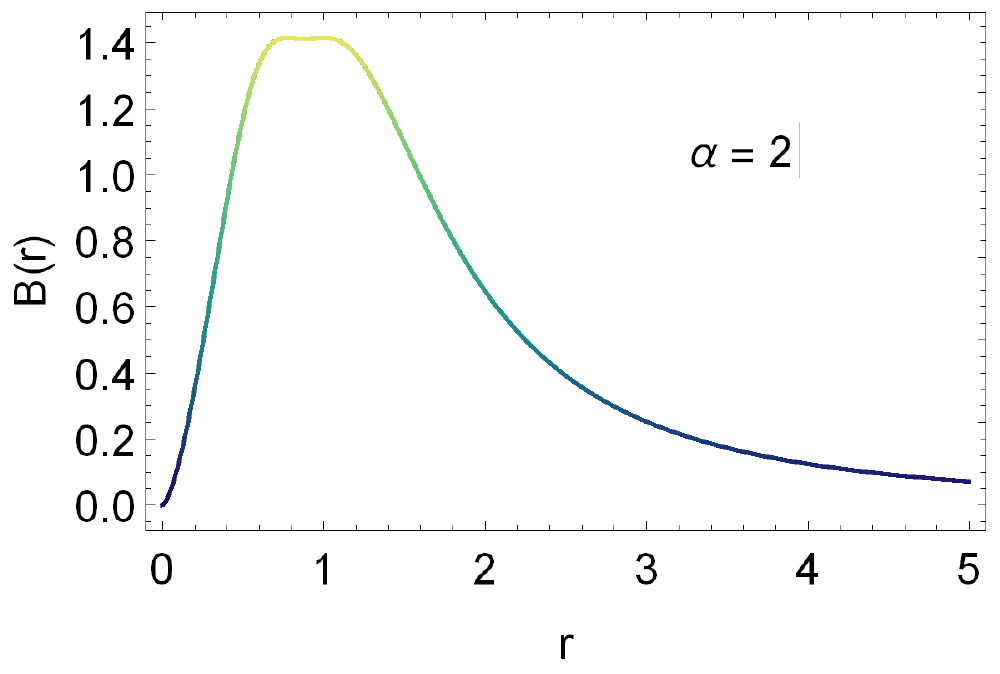}
    \includegraphics[height=3cm,width=3cm]{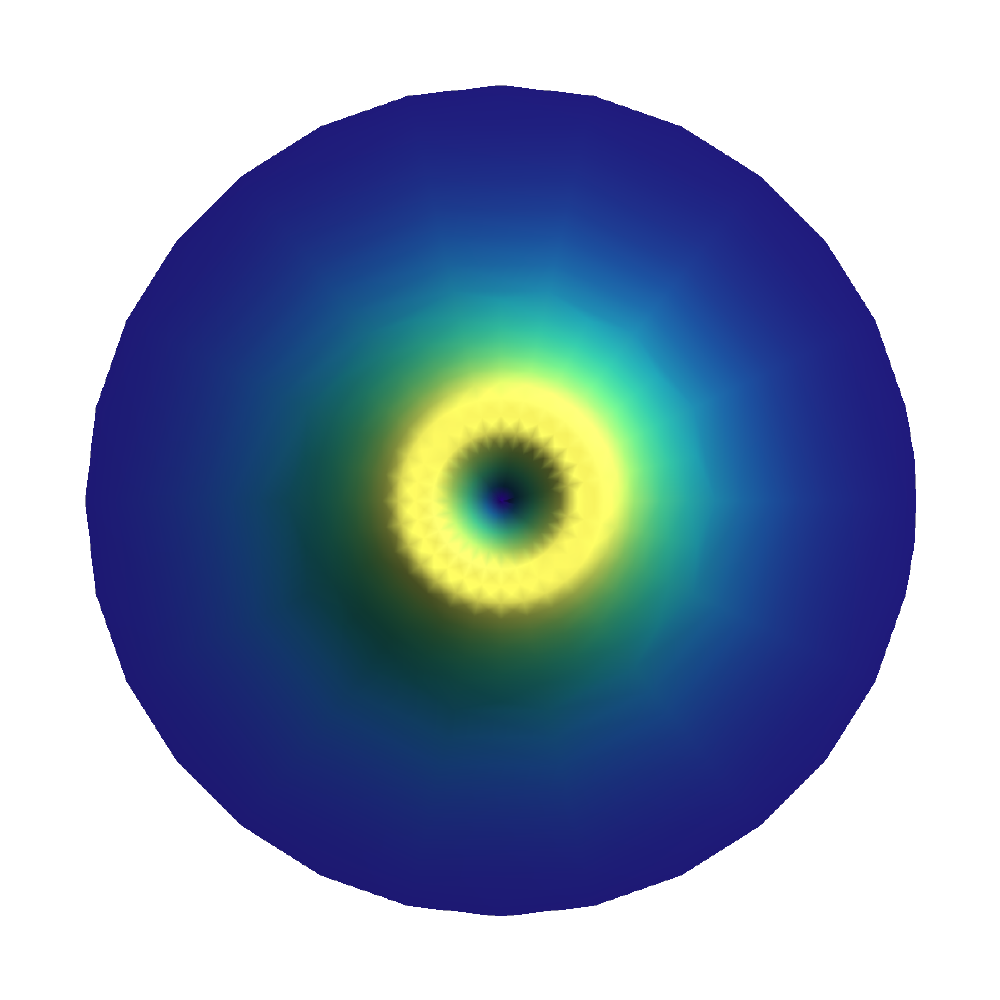}\\
    \includegraphics[height=3.2cm,width=4.2cm]{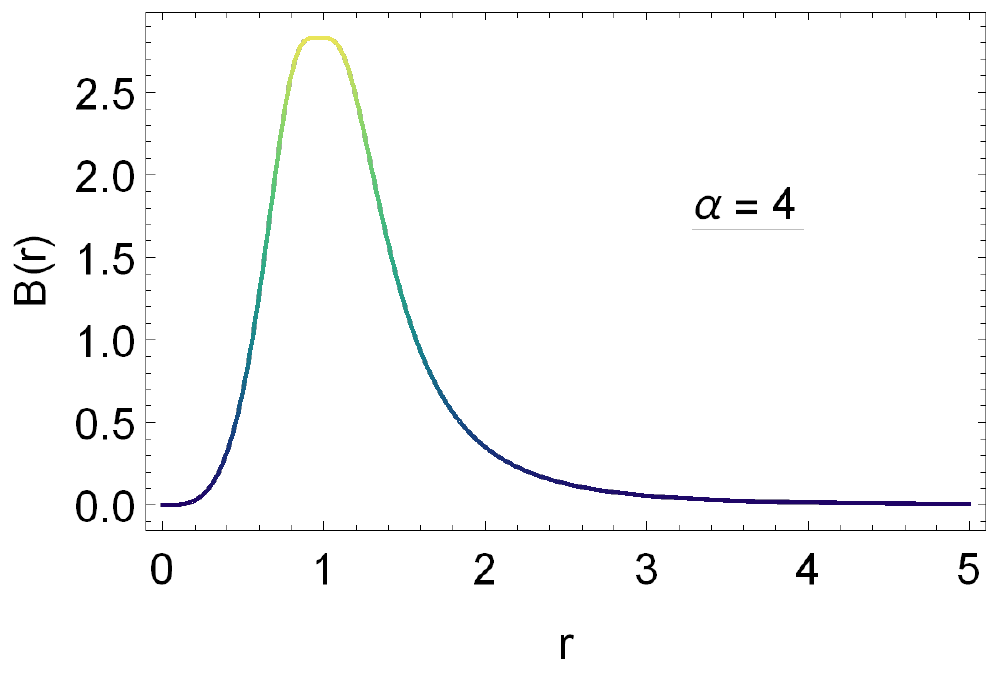}
    \includegraphics[height=3cm,width=3cm]{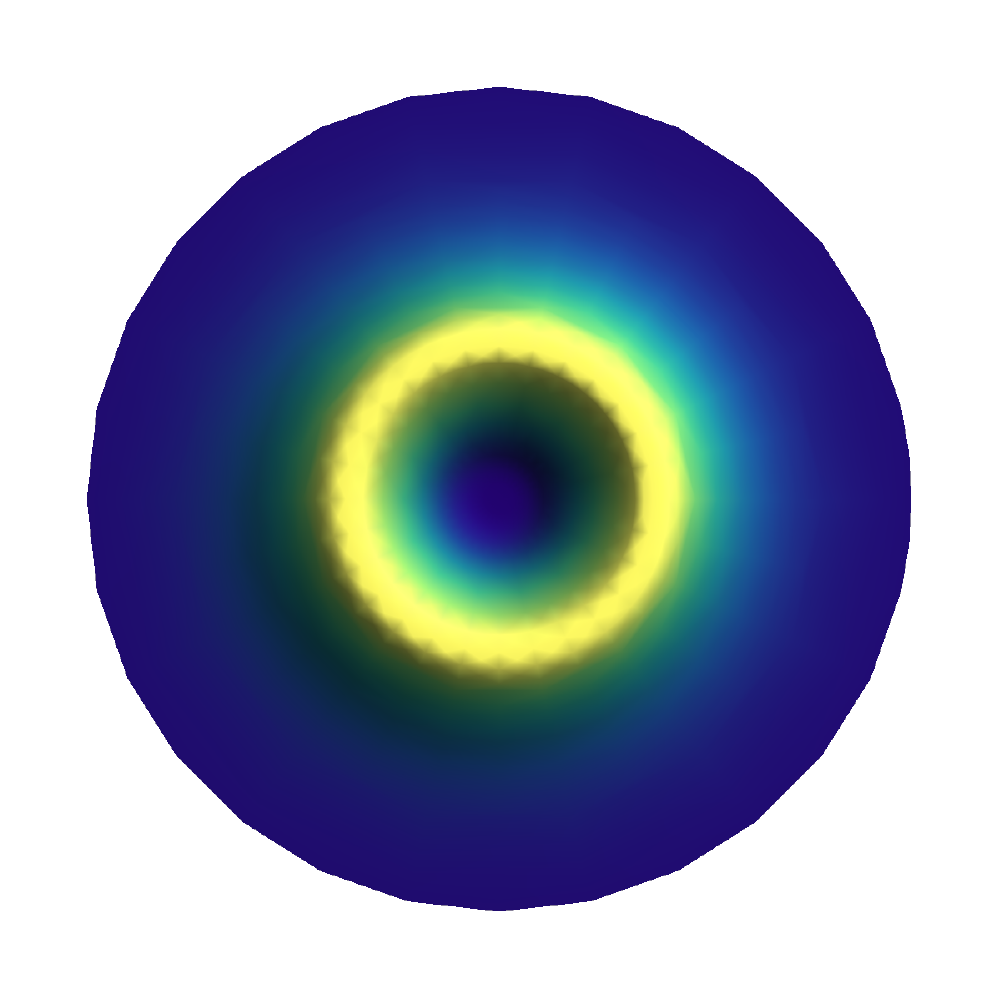}\\
    \includegraphics[height=3.2cm,width=4.2cm]{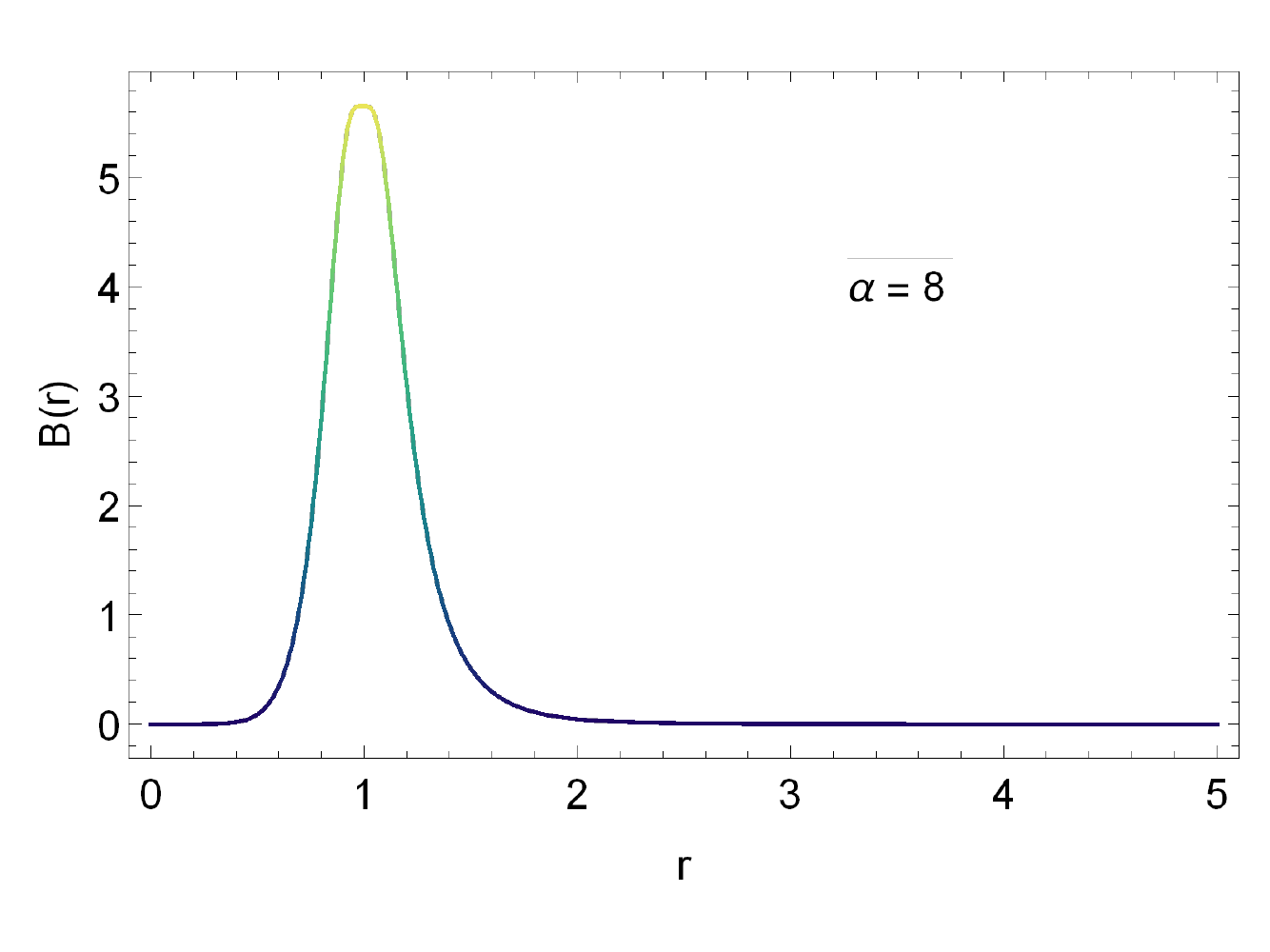}
    \includegraphics[height=3cm,width=3cm]{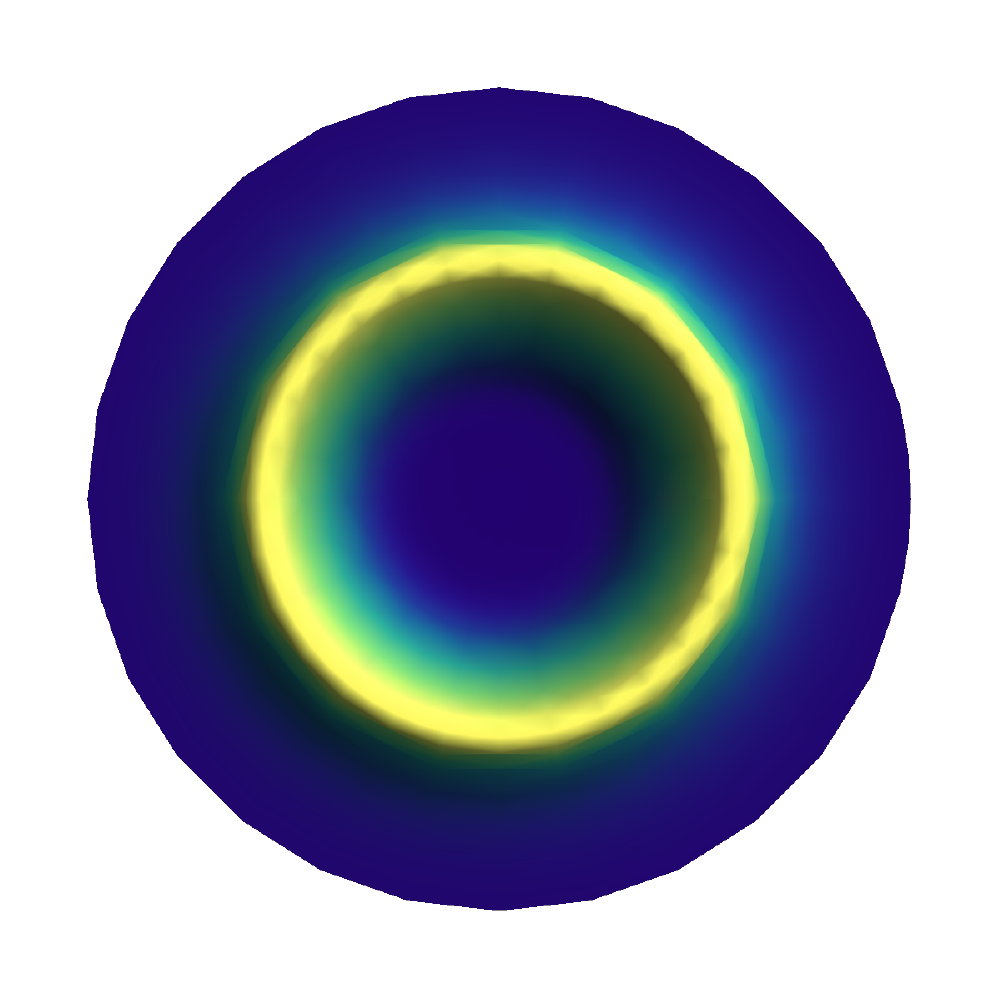}\\
    \includegraphics[height=3.2cm,width=4.2cm]{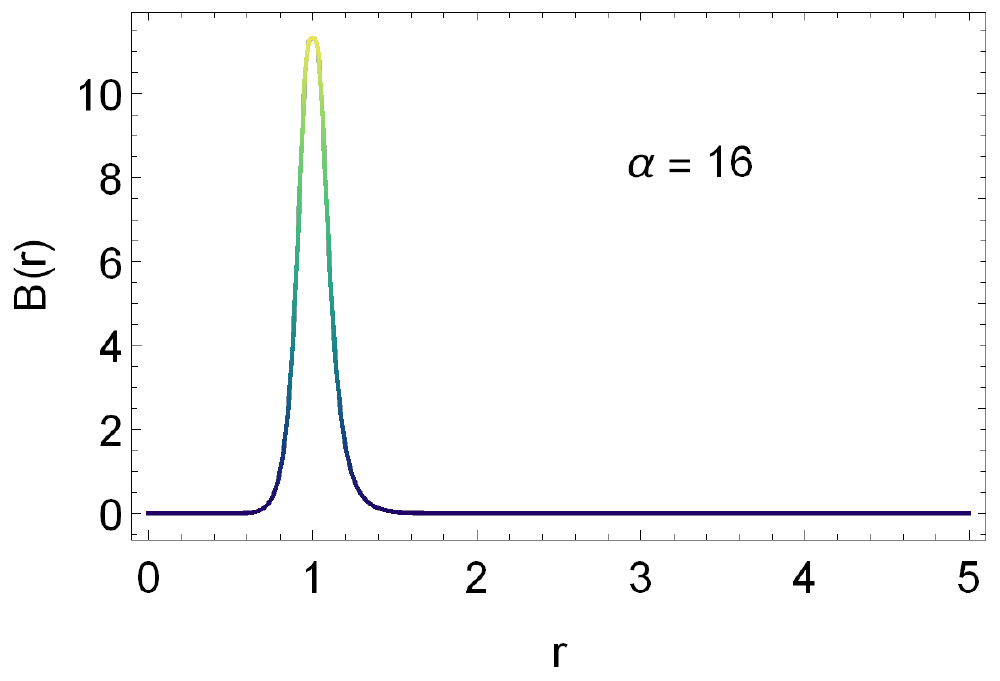}
    \includegraphics[height=3cm,width=3cm]{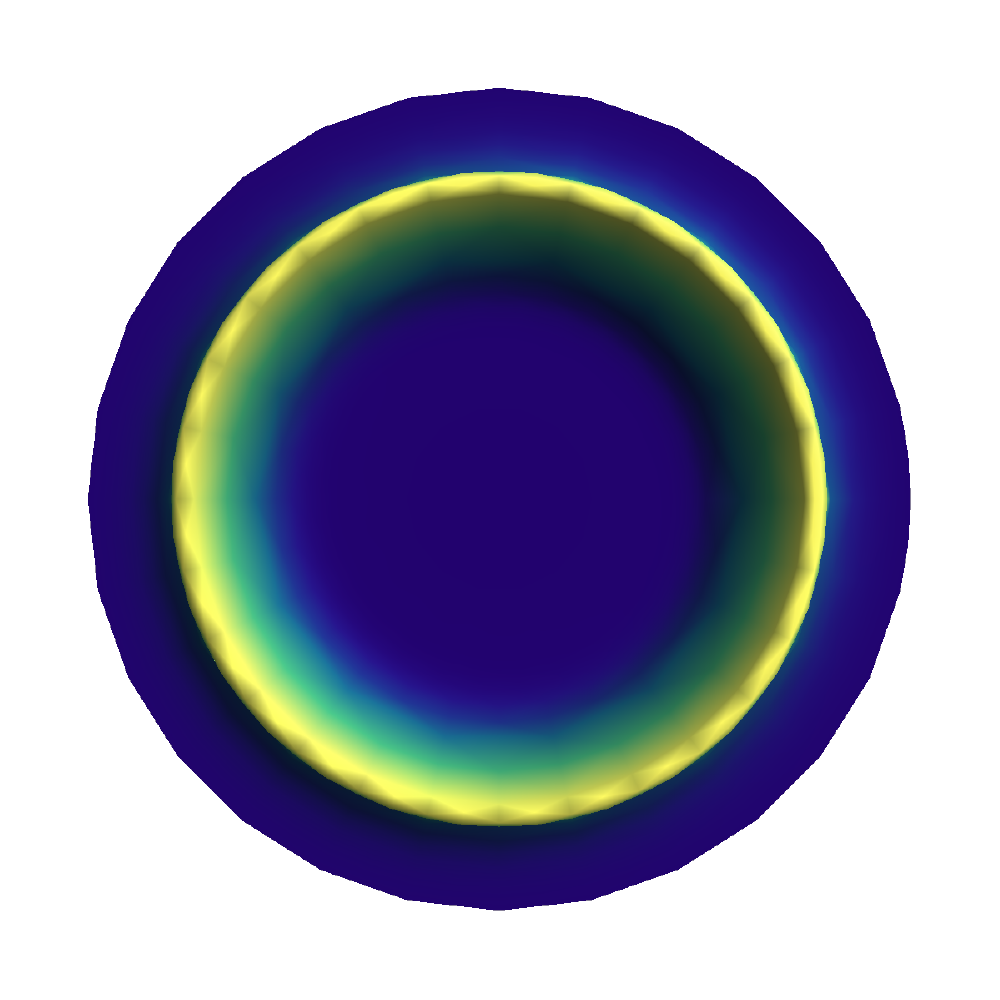}
    \vspace{-0.4cm}
    \caption{Magnetic field varying $\alpha$.}
        \label{fig3}
\end{figure}

Using the numerical solutions of the matter field and the gauge field, one can analyze the magnetic field and the BPS energy density (\ref{BPSED}) of the vortex. Let us start our analysis by investigating the vortex magnetic field. To perform this analysis, we recall that Eq. (\ref{MagneticF}) gives us the magnetic field. Thus, substituting the numerical solution of the gauge field in Eq. (\ref{MagneticF}), we obtain the vortex magnetic field. We expose the magnetic field in Fig. \ref{fig3}. This result shows us an interesting property of the vortex, i. e., the ring-like magnetic field. This feature is what we call a ring-like vortex. For more details, see Ref. \cite{LA}. We discuss more physical implications of these results in the final remarks.

\begin{figure}[ht!]
    \centering
    \includegraphics[height=3.2cm,width=4.2cm]{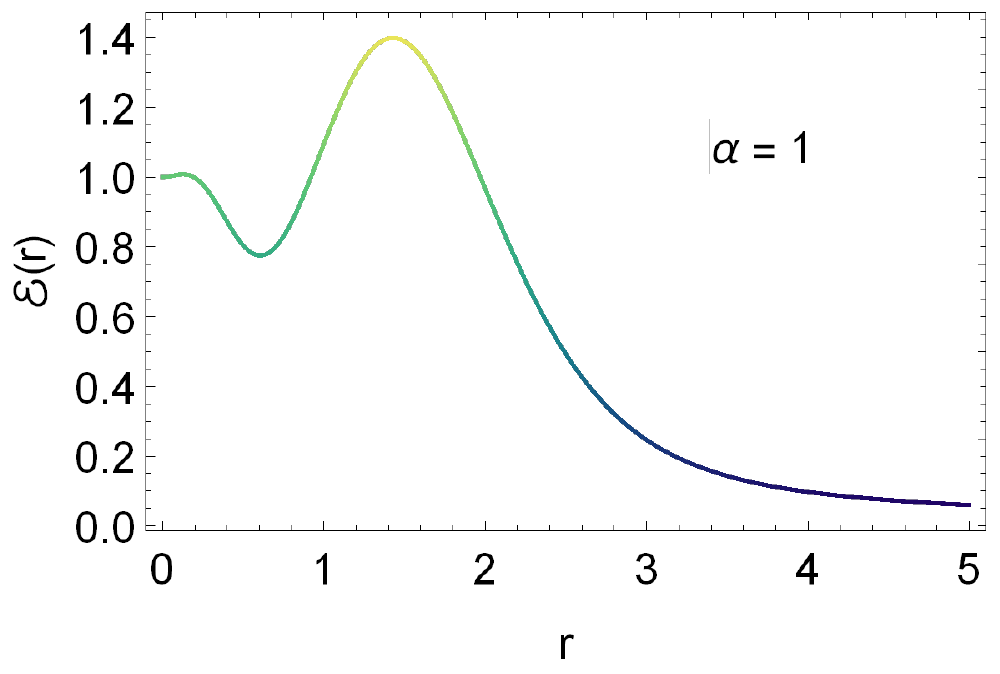}
    \includegraphics[height=3cm,width=3cm]{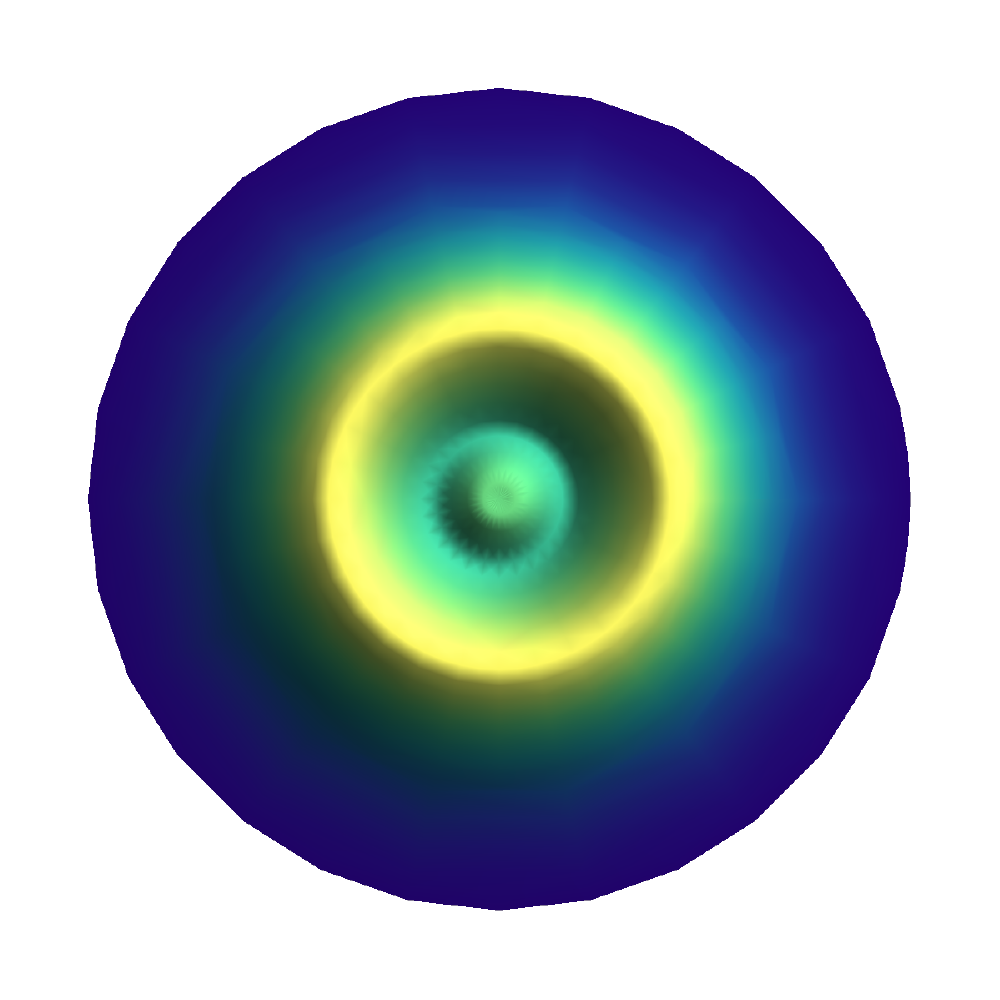}\\
    \includegraphics[height=3.2cm,width=4.2cm]{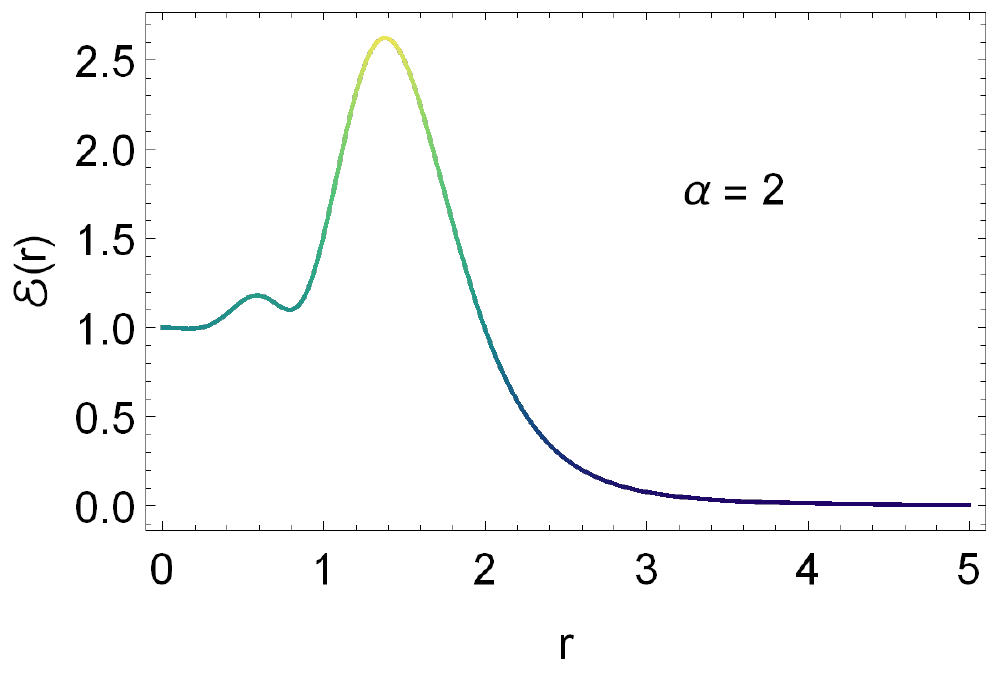}
    \includegraphics[height=3cm,width=3cm]{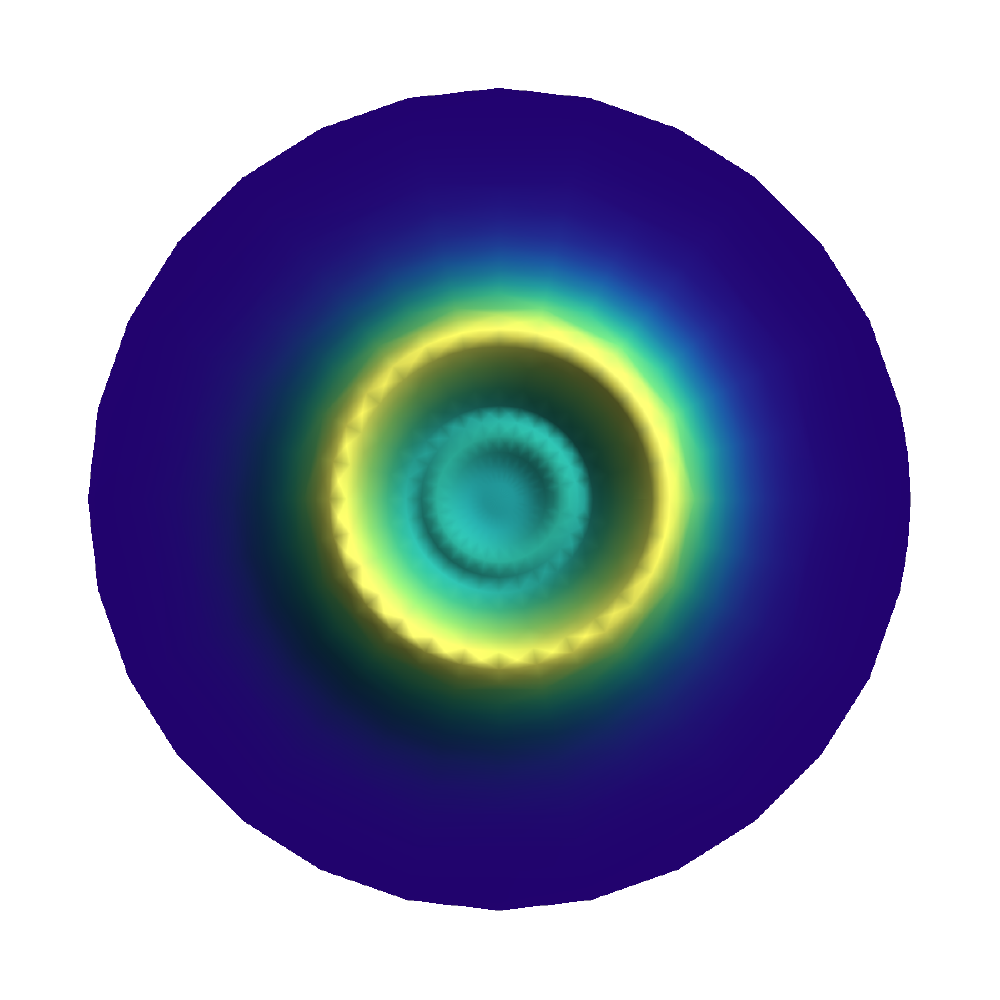}\\
    \includegraphics[height=3.2cm,width=4.2cm]{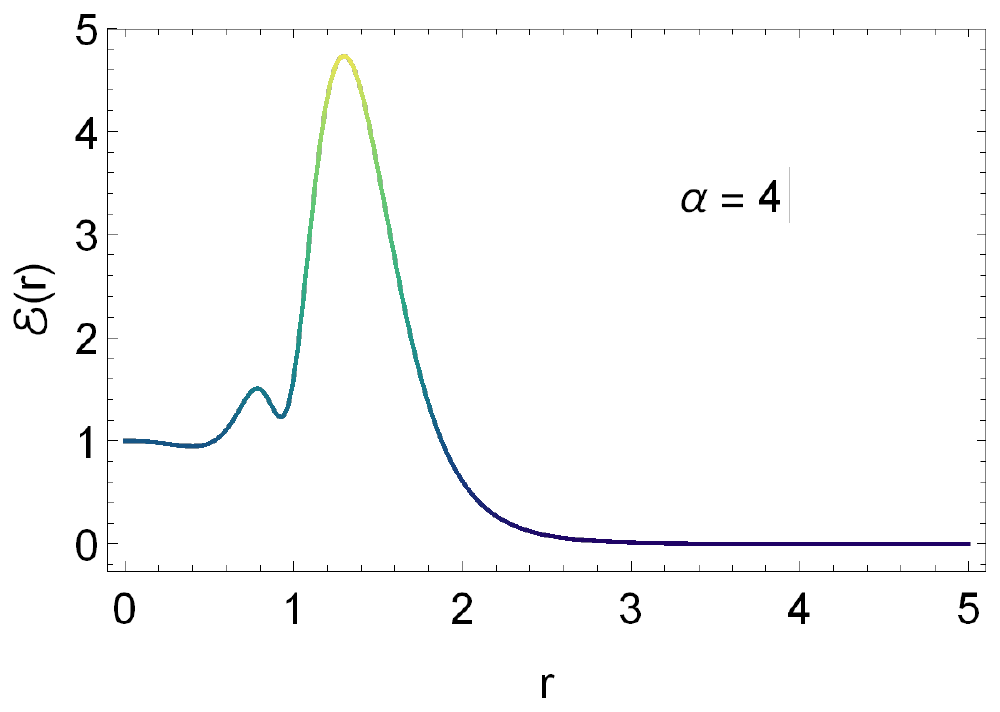}
    \includegraphics[height=3cm,width=3cm]{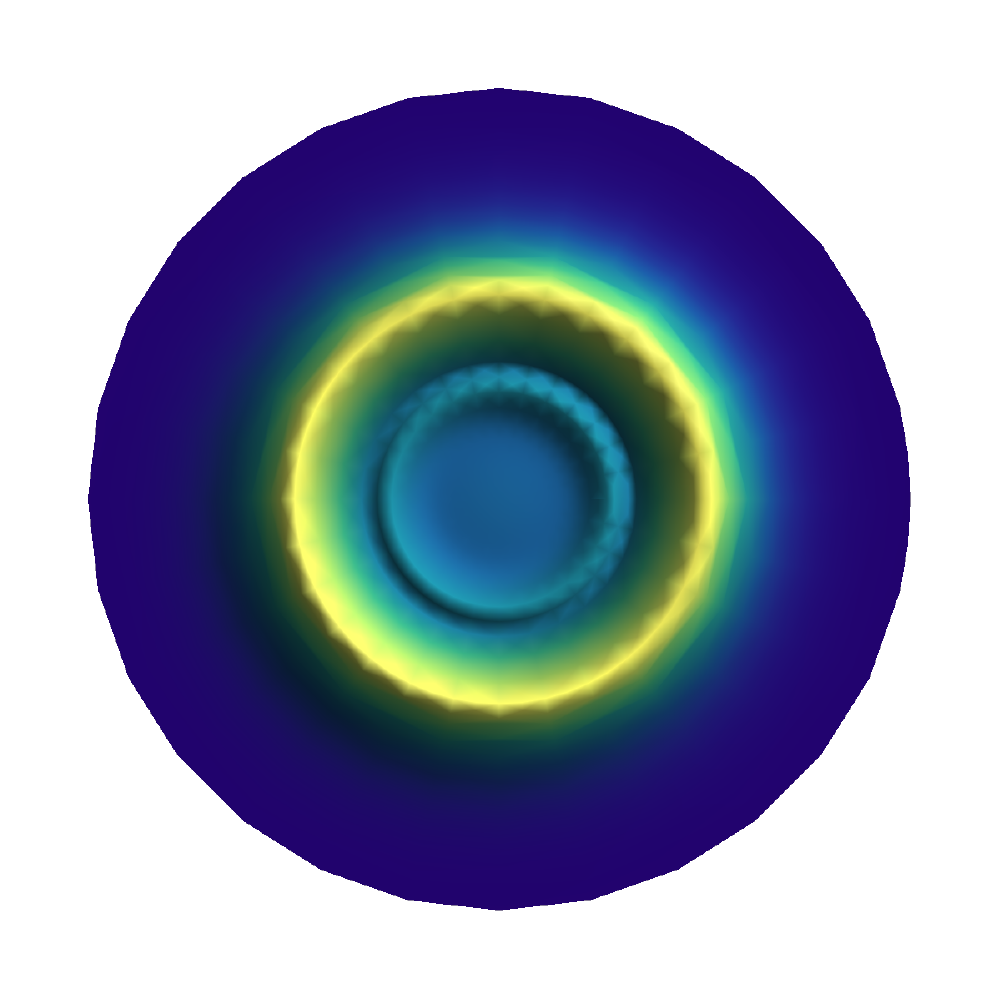}\\
    \includegraphics[height=3.2cm,width=4.2cm]{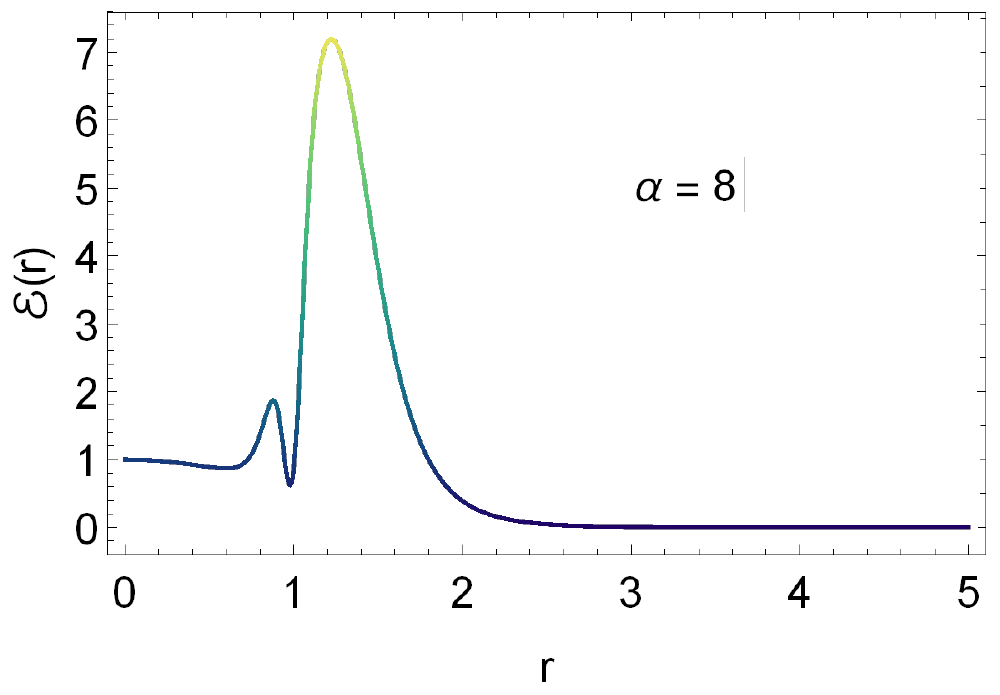}
    \includegraphics[height=3cm,width=3cm]{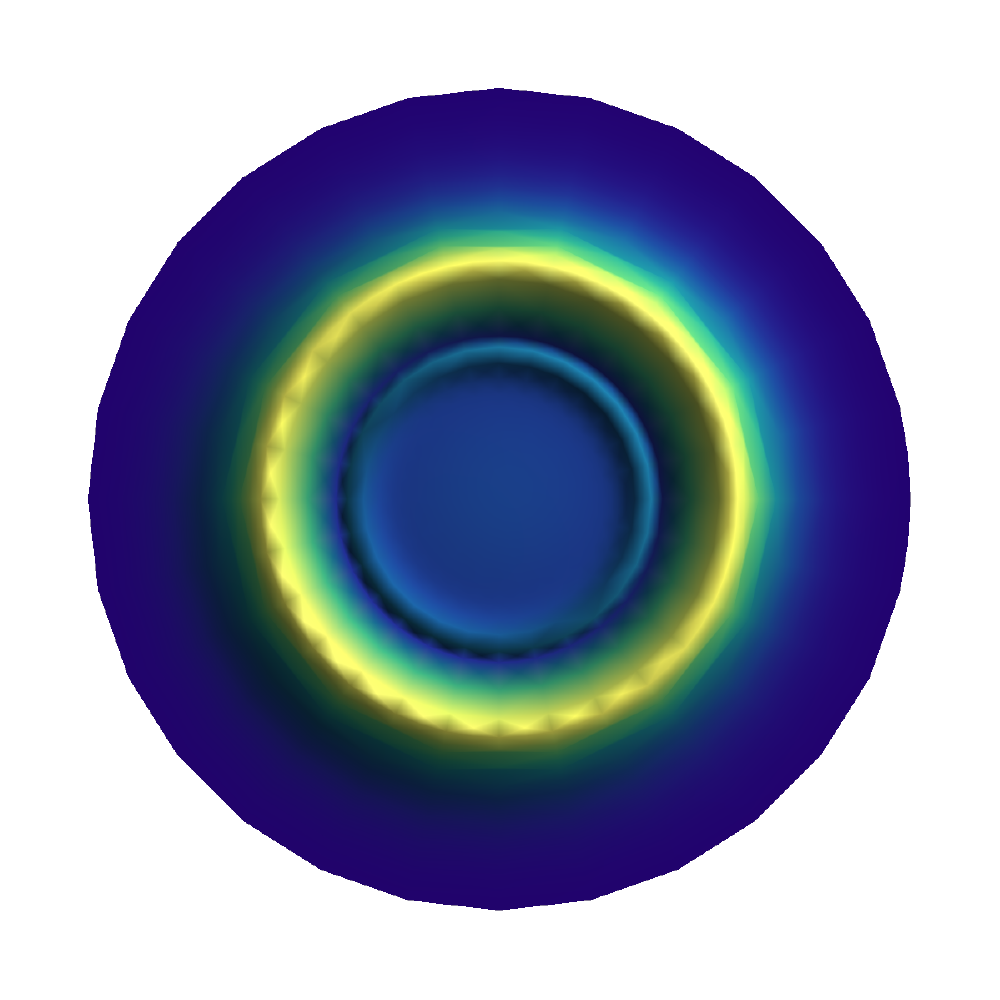}\\
    \includegraphics[height=3.2cm,width=4.2cm]{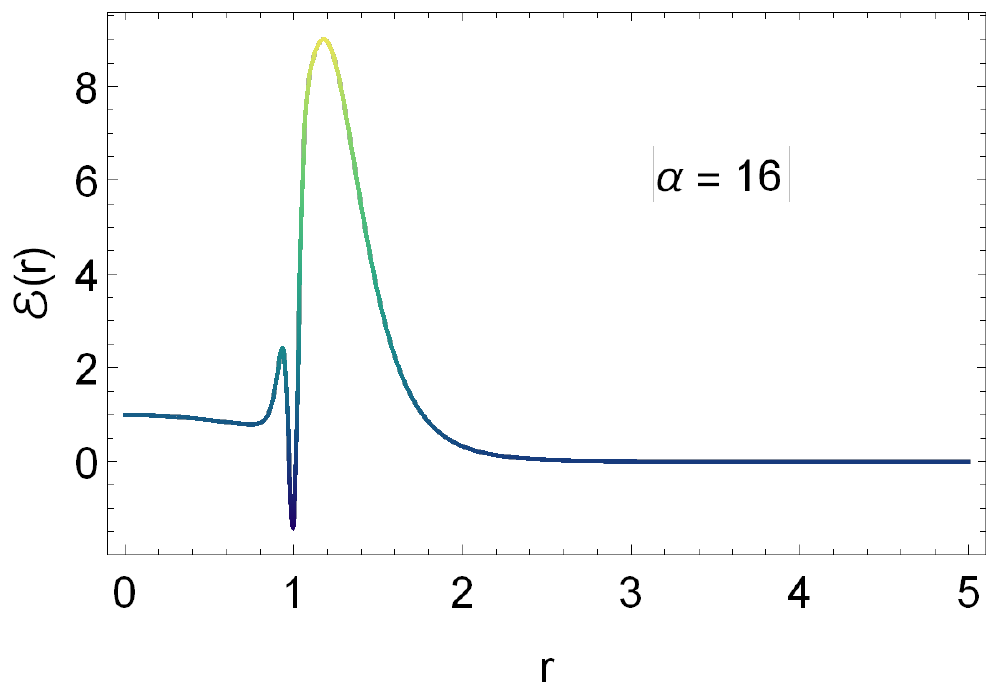}
    \includegraphics[height=3cm,width=3cm]{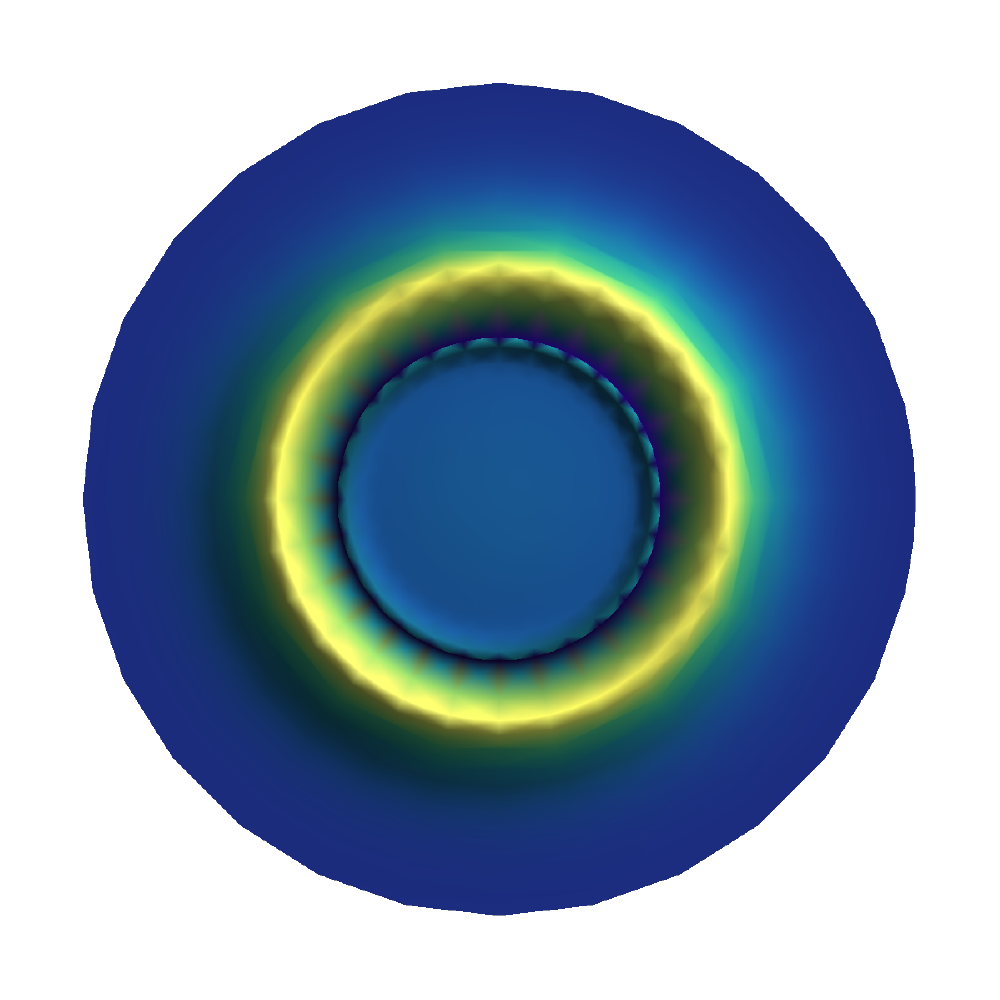}
    \vspace{-0.4cm}
    \caption{BPS energy density varying $\alpha$.}
    \label{fig4}
\end{figure}

By Eq. (\ref{BPSED}), the BPS energy density in terms of the field variable is
\begin{align}\label{DenergyBPS}
    \mathcal{E}_{\text{BPS}}(r)=\pm\frac{1}{r}\frac{d}{dr}\bigg\{N(a(r)-1)\cos f(r)\pm\alpha\tanh[\text{ln}(r^\alpha)]\bigg[1-\frac{1}{3}\tanh^2[\text{ln}(r^\alpha)]\bigg]\bigg\}.
\end{align}
Thus, substituting the numerical solutions of Eqs. (\ref{B1}) in Eq. (\ref{DenergyBPS}), the BPS energy density of the structure is obtained. The Fig. \ref{fig4} shows the numerical solution of the BPS energy density. Analyzing the BPS energy density (see Fig. \ref{fig4}), we highlight the interesting appearance of internal structures.

0
Naturally, one notes by analyzing the magnetic field and the vortex energy that the magnetic field engenders ring-like profiles when the $\alpha$ parameter increases, i.e., when the contribution of the cuscuton-like non-canonical term increases. Physically, these rings appear in theory when the vacuum from the $\psi(r)$ sector, and consequently of the sigma sector, becomes more localized. While this phenomenon occurs, the energy density supports an almost constant profile around a ring-like internal structure. This behavior informs us that physically, at the vortex core, the matter seems unable to generate magnetic fields when the non-canonical contributions become more significant, i.e., the vacuum states are less dispersed. Generally speaking, this structure is similar to what happens in superconducting baryonic tubes in the low-energy limit of QCD, where these structures describe superconducting baryonic tubes with the baryonic charge and energy concentrated inside structures \cite{Can0}. In this sense, the vortices found in this work are similar to baryonic superconductor tubes that carry a persistent current (which disappears outside the vortex).

\section{Final remarks}

In this work, we studied the vortex solutions of a multi-field theory. The model proposed has a canonical field, i. e., the field describing the O(3)-sigma model, and a non-canonical field, i. e., the field $\psi$. Furthermore, it is considered that $\Phi$ is non-minimally coupled with the gauge field. Thus, the vortices produced have an anomalous contribution from the magnetic dipole momentum.

We consider that the scalar field dynamics have canonical and non-canonical contributions. These contributions are, respectively, $\frac{1}{2}\partial_\mu\psi\partial^\mu\psi$  and $\eta\sqrt {\vert\partial_\mu\psi\partial^\mu\psi\vert}$. The non-canonical contribution is what is known as cuscuton. The cuscuton term is interesting since the contribution of the cuscuton in the stationary case is trivial. Thus, the equation of motion will only have contributions from the canonical terms in the stationary limit. However, in this case, the term cuscuton will have a non-trivial contribution to the energy density of the structures. Therefore, in the stationary BPS limit, cuscuton will be an impurity of the theory. It is worthwhile to mention that cuscuton, in this scenario, is interpreted as an impurity only at the topological sector of the sigma field. Indeed, this is a consequence of dealing with a vacuumless theory, i. e., $\mathcal{V}=0$.

Furthermore, the vacuumless multi-field model proposed proved to support electrically neutral vortices that engender an interesting internal structure. Besides, the magnetic field of vortices also has a ring-like shape. Note that these ring structures become well defined if the contribution of cuscuton increases, i. e. when the $\alpha$ parameter increases. Consequently, as $\eta$ increases, the flux of the magnetic field will increase, and therefore, the energy radiated by the vortex increases. In general, we can interpret this as a consequence of the behavior of the matter field and the gauge field in the topological sector of the sigma model. These fields have a very peculiar behavior, i. e., when the contribution of the cuscuton term (the impurity) increases, the matter field and the gauge field become more compact. That occurs due to the location of the kink at the topological sector of $\psi$ around $r=1$.

Finally, allow us to mention that theories of supersymmetric vortices are a subject of growing interest. That is because these theories generalize particle-vortex dualities. Thus, one expects that duality to have applications in condensed matter physics. Therefore, a future perspective of this work is the study of particle-vortex duality in our theory. Furthermore, one can build extensions of this theory by implementing these structures in dielectric media. We hope to carry out these studies soon.

\section{Acknowledgment}

F. C. E. Lima (FCEL) expresses gratitude to the Coordenação de Aperfeiçoamento do Pessoal de Nível Superior (CAPES) for the doctoral scholarship number 88887.372425/2019-00 received from August 2019 to July 2023. C. A. S. Almeida (CASA) is thankful to the Conselho Nacional de Desenvolvimento Científico e Tecnológico (CNPq), grant number 88887.372425/2019-00, for financial support. Additionally, CASA would like to thank the Print-UFC CAPES program, project number 23067.009493/2023-79, for funding. C. A. S. Almeida acknowledges the Department of Physics and Astronomy at Tufts University for its warm hospitality. The authors thank the anonymous referee for their criticism, comments, and suggestions.


\section*{Data Availability}
The datasets generated during and/or analysed during the current study are available from the corresponding author on reasonable request.

\section*{Appendix A - The analytical solution of the $\psi(r)$ field}

Considering Eq. (\ref{PsiE}), we can easily demonstrate its analytical solution [presented in Eq. (\ref{tanhln})], rewriting Eq. (\ref{PsiE}) as follows
\begin{align}
    \int \frac{d\psi}{1-\psi^2}=\pm\alpha\int \frac{dr}{r}.
\end{align}

Note that assuming $\tanh(\psi)=u$, one obtains
\begin{align}
    \int\frac{d\psi}{1-\psi^2}\equiv \int du=u\,\to\, \int\frac{d\psi}{1-\psi^2}=\text{arctanh}(\psi)+c_0.
\end{align}
where $c_0$ is an integration constant. Thus, Eq. (\ref{PsiE}), leads us to
\begin{align}
    \text{arctanh}(\psi)=\pm \text{ln}(r^{\alpha})\pm \lambda_0,
\end{align}
where $\lambda_0$ is a new integration constant. The expression above leads us to
\begin{align}
    \psi(r)=\pm\text{tanh}[\text{ln}(r^\alpha)+\lambda_0].
\end{align}
For the topological conditions of the $\psi(r)$ field to be respected, i.e., $\psi(r\to -\infty)=\mp 1$ and $\psi(r\to \infty)=\pm 1$, one can adopt $\lambda_0=0$, once this condition does not modify the dynamics of the field. Therefore,
\begin{align}
    \psi(r)=\pm\text{tanh}[\text{ln}(r^{\alpha})].
\end{align}
\end{document}